\documentclass{PoS}
\usepackage{subfigure}
\usepackage{wrapfig}
\usepackage{cite}
\usepackage{url}

\title{Progress on the nature of the QCD thermal transition as a
   function of quark flavors and masses}

\ShortTitle{The nature of the QCD thermal transition as a
   function of quark flavors and masses}

\author{\speaker{Francesca Cuteri}\\%\thanks{A footnote may follow.}\\
        Institut f\"ur Theoretische Physik - Goethe-Universit\"at Frankfurt am Main\\
        Max-von-Laue-Str. 1, 60438 Frankfurt am Main\\
        E-mail: \email{cuteri@th.physik.uni-frankfurt.de}}
\author{Owe Philipsen\\
        Institut f\"ur Theoretische Physik - Goethe-Universit\"at Frankfurt am Main\\
        Max-von-Laue-Str. 1, 60438 Frankfurt am Main,\\
        John von Neumann Institute for Computing (NIC) GSI\\
        Planckstr. 1, 64291 Darmstadt, Germany\\
        E-mail: \email{philipsen@th.physik.uni-frankfurt.de}}
\author{Alessandro Sciarra\\
        Institut f\"ur Theoretische Physik - Goethe-Universit\"at Frankfurt am Main\\
        Max-von-Laue-Str. 1, 60438 Frankfurt am Main\\
        E-mail: \email{sciarra@th.physik.uni-frankfurt.de}}

\usepackage{amsmath}

\newcommand{\NSigma}{N_\sigma}
\newcommand{\NTau}{N_\tau}
\newcommand{\Nf}{N_\text{f}}

\newcommand{\Loewe}{LOEWE-CSC}
\newcommand{\Lcsc}{L-CSC}
\newcommand{\clqcd}{CL\kern-.25em\textsuperscript{2}QCD}
\newcommand{\Ocl}{OpenCL}

\newcommand{\psibar}{\bar{\psi}}
\newcommand{\chiralcond}{\langle \psibar \psi \rangle}
\newcommand{\ms}{m_{s}}
\newcommand{\mud}{m_{u,d}}
\newcommand{\Action}{\mathcal S}
\newcommand{\SGluon}{\Action_{\text{G}}}

\newcommand{\Binder}{B_4}
\newcommand{\Skewness}{B_3}
\newcommand{\ZTwoUniversality}{Z_2}
\newcommand{\NfTricr}{\Nf^\text{tric}}
\newcommand{\NfLin}{\Nf^\text{lin}}

\newcommand{\Link}{U}

\newcommand{\LatMassStaggered}{m}
\newcommand{\LatMassStaggeredZTwo}{\LatMassStaggered_{\ZTwoUniversality}}

\newcommand{\LatCoupling}{\beta}

\abstract{We investigate to which extent we can exploit the dependence of the order of the chiral transition on the number of light degenerate flavors $\Nf$, re-interpreted as continuous parameter in the path integral formulation, as a means to perform a controlled chiral extrapolation and deduce the order of the transition for the case $\Nf=2$, which is still under debate.}

\FullConference{The 36th Annual International Symposium on Lattice Field Theory - LATTICE2018\\
		22-28 July, 2018\\
		Michigan State University, East Lansing, Michigan, USA.}

\begin{document}

\section{Introduction}
\label{sec:introduction}

\begin{figure}[tp]
   \centering
   \subfigure[First order scenario in the $\ms-\mud$ plane]%
             {\label{fig:firstOrderScenarioCP}\includegraphics[width=0.45\textwidth,clip]{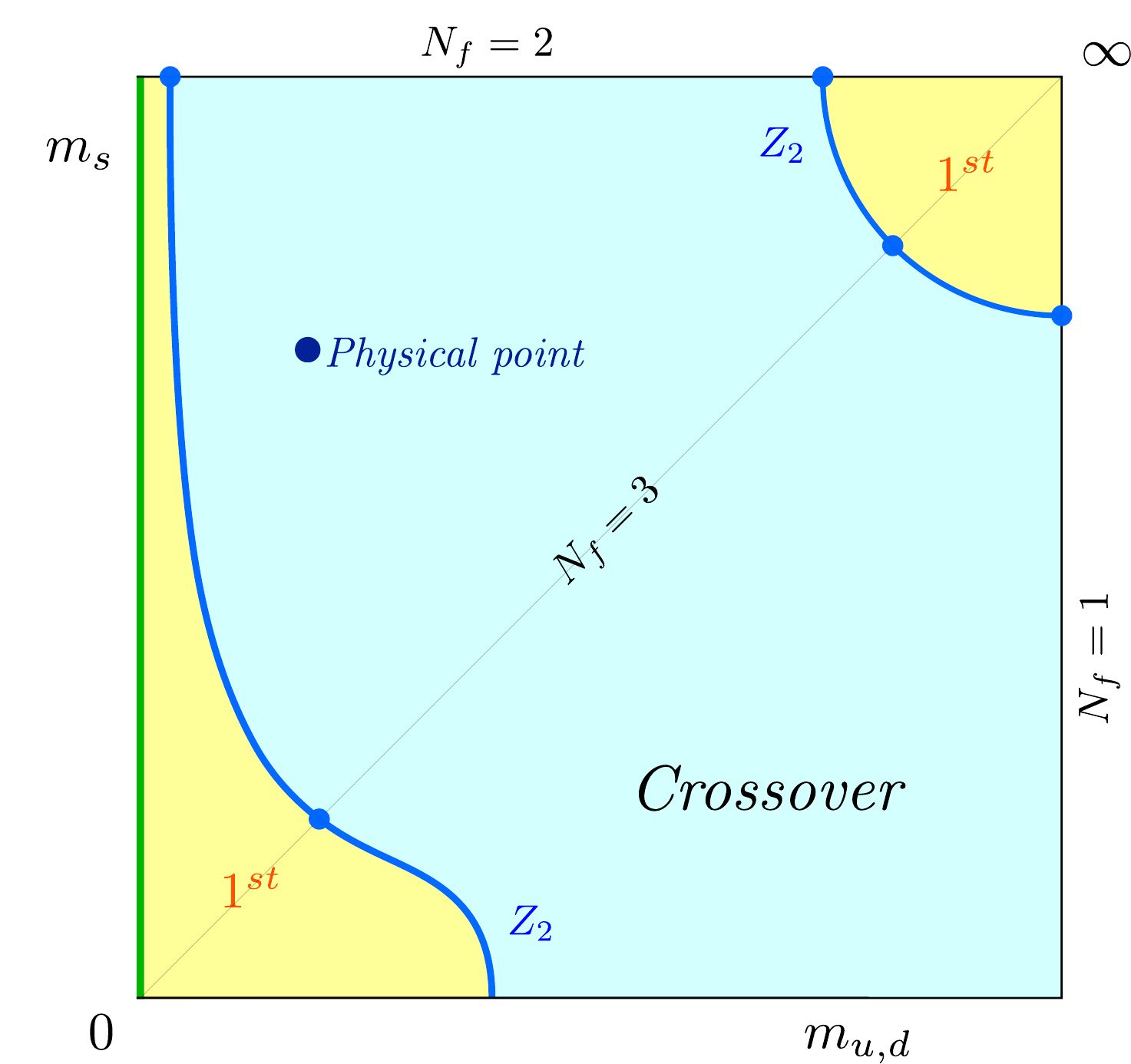}}\hfill
   \subfigure[Second order scenario in the $\ms-\mud$ plane.]%
             {\label{fig:secondOrderScenarioCP}\includegraphics[width=0.45\textwidth,clip]{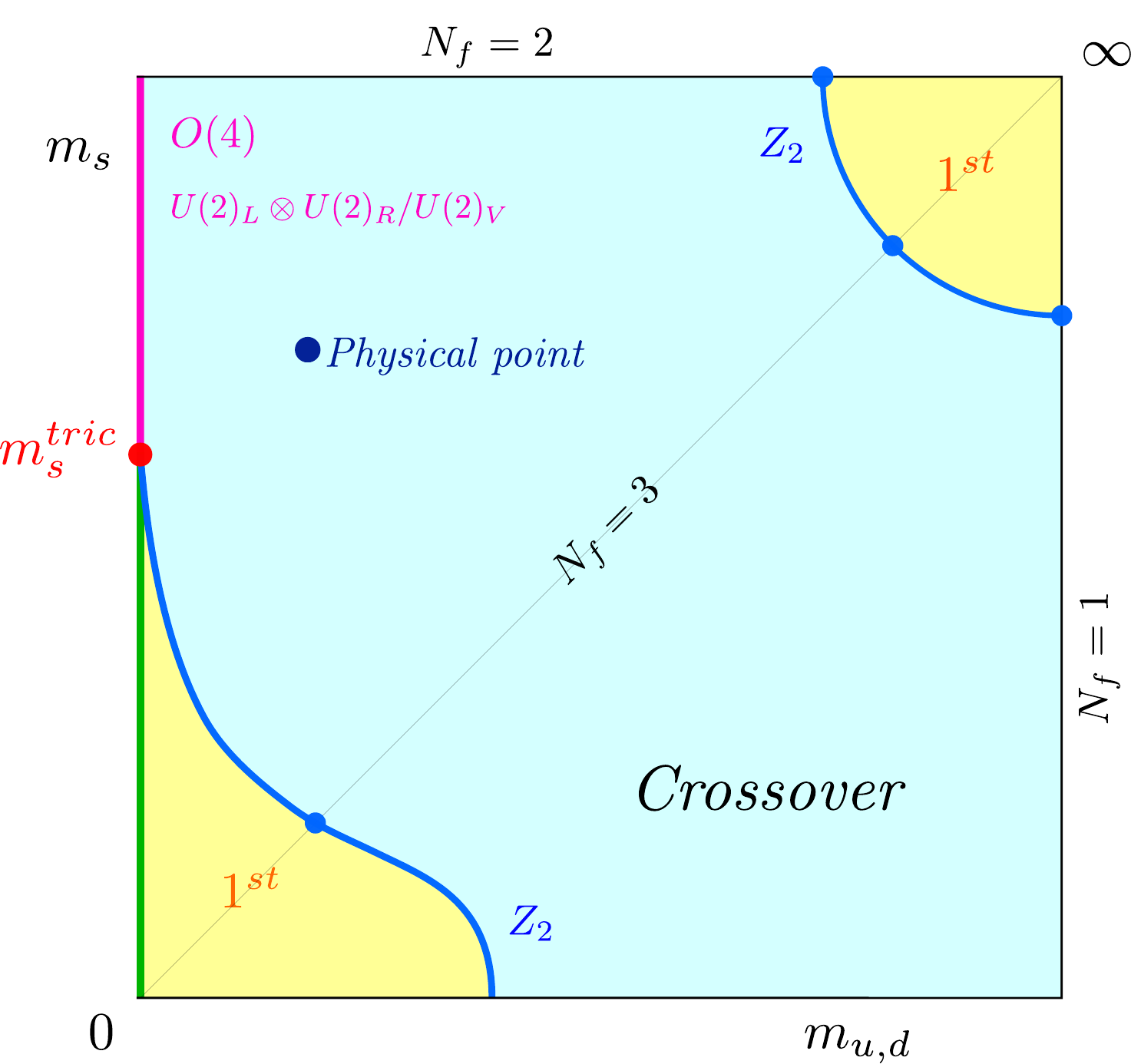}}
   \caption{Two possible scenarios for the order of the QCD thermal phase transition as function of the masses of quarks. Indicated in Fig.~\protect\ref{fig:secondOrderScenarioCP} are also plausible universality classes for the second order line at $\mud=0$.}
   \label{fig:scenariosCP}
\end{figure}

The \emph{Columbia plot}, of which we show in \figurename~\ref{fig:scenariosCP} two possible versions based on current findings, encapsulates our still very limited knowledge about the order of the thermal phase transition in QCD as function of the two light (assumed degenerate) quark masses $\mud$ and the strange quark mass $\ms$. Continuum extrapolated results are so far only available at the physical point. Elsewhere, using different unimproved~\cite{Karsch:2001nf,deForcrand:2003vyj,deForcrand:2007rq,Bonati:2014kpa,Philipsen:2016hkv,deForcrand:2017cgb} and improved~\cite{Burger:2011zc,Brandt:2016daq,Jin:2017jjp,Bazavov:2017xul} fermion discretizations, seemingly contradicting results have been obtained, in particular in what concerns the case of $\Nf=2,3$ degenerate light flavors in the limit of small masses corresponding to the top and bottom left corners in the Columbia plot, respectively.

\begin{figure}[tp]
   \centering
   \subfigure[First order scenario in the $\LatMassStaggered-\Nf$ plane]%
             {\label{fig:firstOrderScenarioMvsNf}\includegraphics[width=0.45\textwidth,clip]{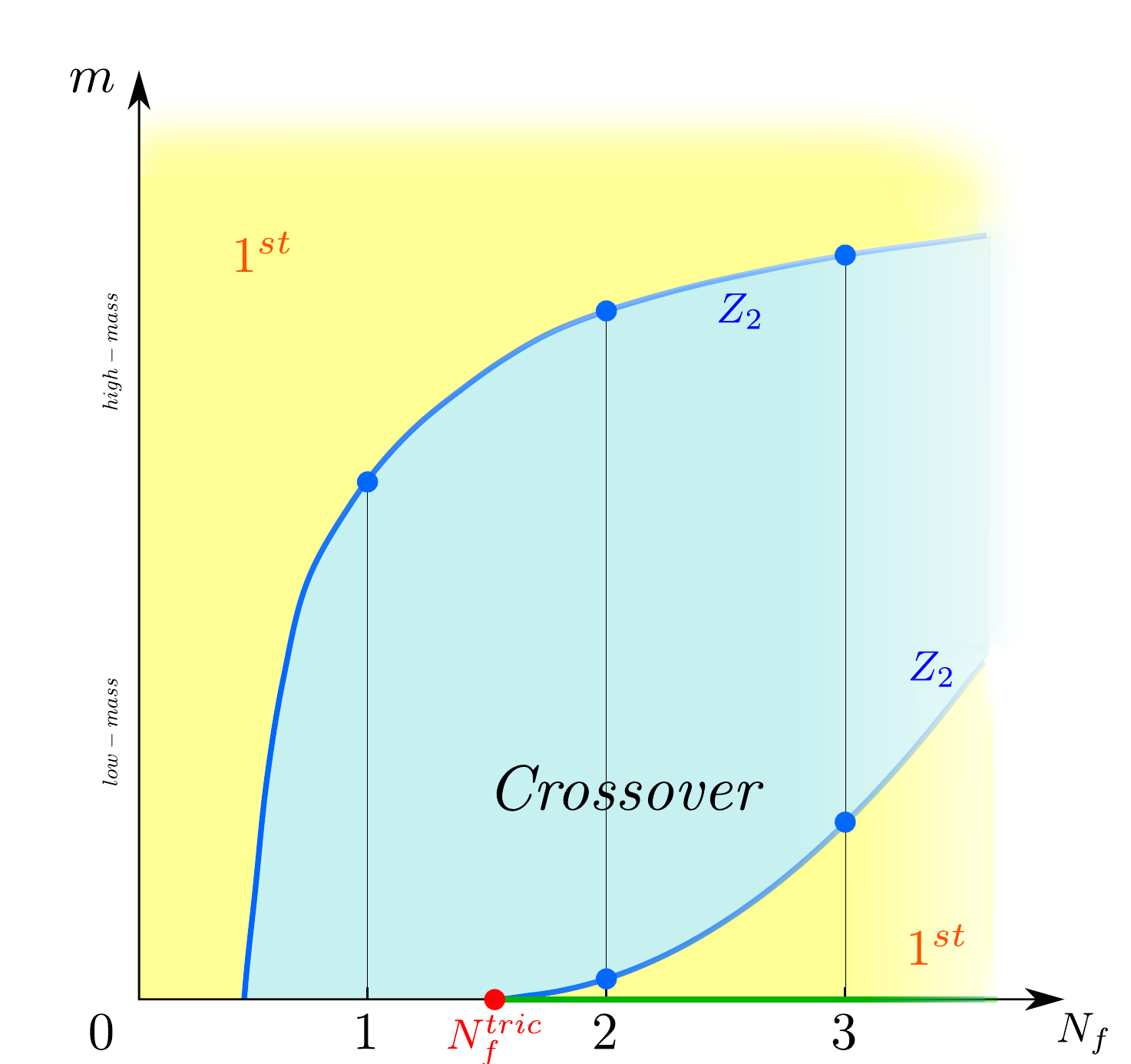}}\hfill
   \subfigure[Second order scenario in the $\LatMassStaggered-\Nf$ plane]%
             {\label{fig:secondOrderScenariosMvsNf}\includegraphics[width=0.45\textwidth,clip]{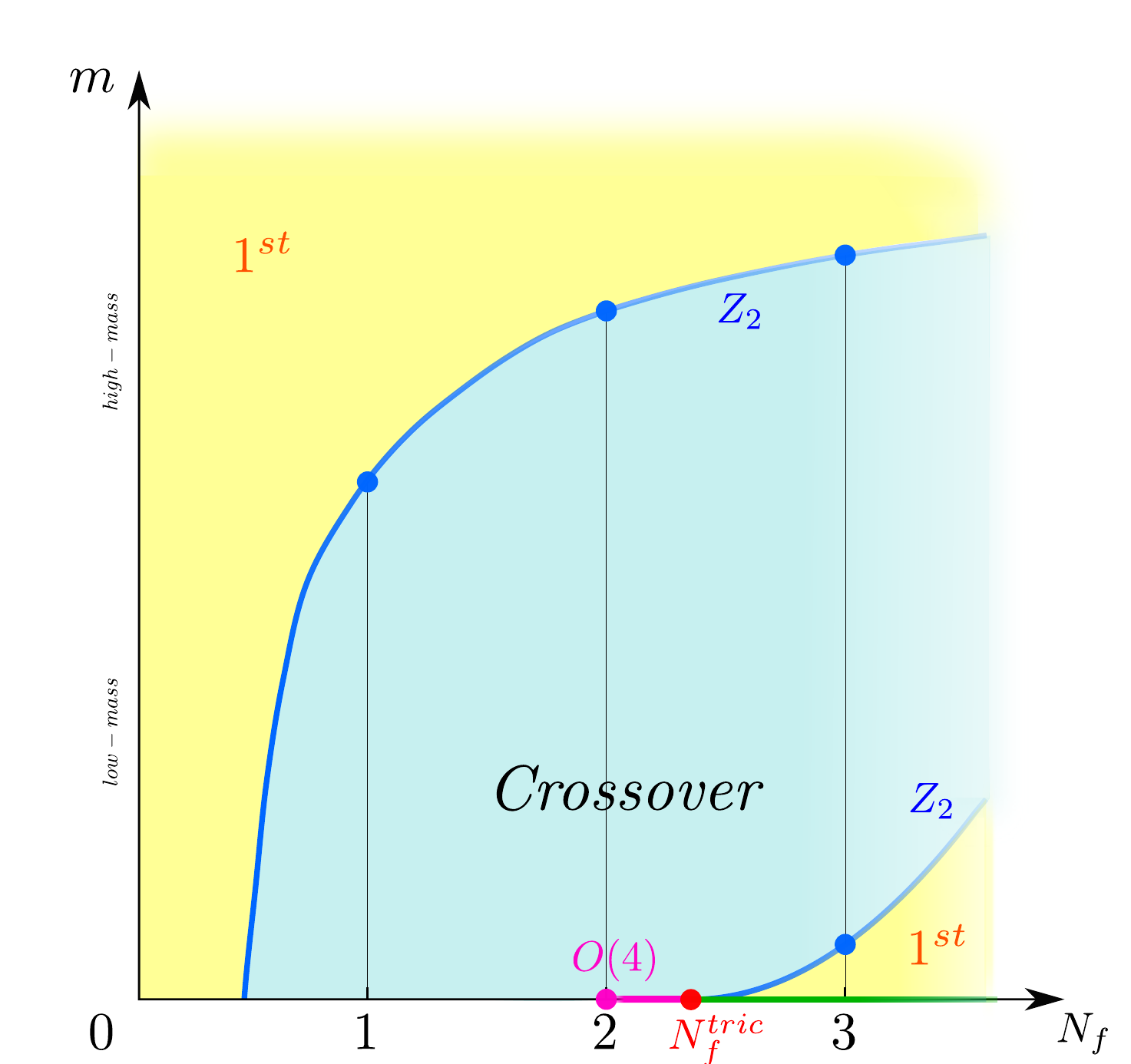}}
   \caption{The two considered possible scenarios for the order of the QCD thermal phase transition as function of the mass of the quarks and the number of degenerate fermion flavors.}
   \label{fig:scenariosMvsNf}
\end{figure}

This motivated us to push forward with studies aiming at elucidating, in particular, the picture for $\Nf=2$ degenerate light flavors, by exploiting the dependence of the chiral transition on the number of light degenerate flavors $\Nf$ as a means to perform controlled chiral extrapolations.
To this end, we treated $\Nf$ as a continuous real parameter, of some statistical system behaving, at any integer $\Nf$ value, as QCD at zero density, with $\Nf$ mass-degenerate fermion species~\cite{Cuteri:2017gci}
\begin{equation}
    Z_{\Nf}(m) = \int \mathcal{D}\Link \left[\det M(\Link,m)\right]^{\Nf} e^{-\SGluon}\;.
\end{equation}
Within this framework, the two considered scenarios for the Columbia plot can be put in one-to-one correspondence with the two sketches for the order of the thermal phase transition in the $(\LatMassStaggered,\Nf)$-plane displayed in Figure~\ref{fig:scenariosMvsNf}.

Our original strategy was to find out for which (tricritical) value $\NfTricr$ the phase transition displayed by this system changes from first-order to second-order, by mapping out the $\ZTwoUniversality$ phase boundary.
The extrapolation to the chiral limit with known tricritical exponents can then decide between the two scenarios, depending on whether $\NfTricr$ is larger or smaller than 2.

While the tricritical scaling region was found to be very narrow already on coarse lattices, results at larger $\LatMassStaggered$ and $\Nf$ were found to feature, over a much wider region, a remarkable linear behavior, which was not expected on universality grounds.

\begin{wrapfigure}{r}{0.5\textwidth}
   \centering
              {\label{fig:rescaled}\includegraphics[width=0.5\textwidth,clip]{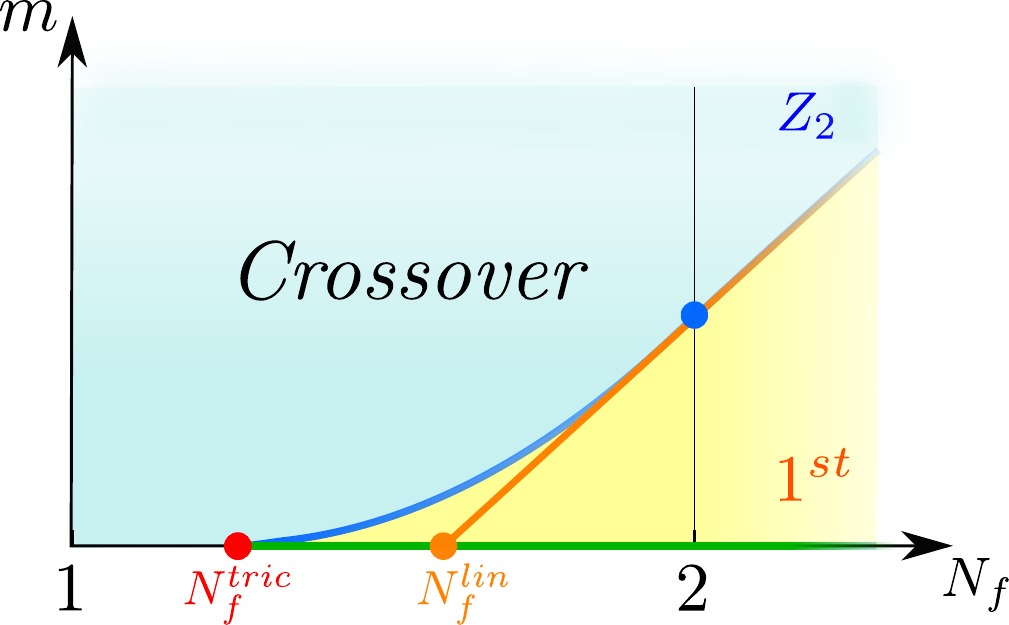}}
   \caption{Sketch showing how, via a linear extrapolation to the chiral limit, $\NfLin$ as an upper bound for $\NfTricr$ can be extracted, with the first order scenario being realized for as long as $\NfLin < 2$.}
\end{wrapfigure}

What our findings suggest is that, if it is reasonable to expect both linearity within some range in $\Nf$ and tricritical scaling more in the chiral limit, then one would be able to make use of a linear extrapolation to $\LatMassStaggered=0$, to at least get $\NfLin$ as an upper bound for $\NfTricr$, out of much more affordable simulations and possibly without even simulating at noninteger numbers of flavors.

For as long as the upper bound from the linear extrapolation keeps lying  at $\Nf < 2$, while one simulates at larger and larger $\NTau$ values towards the continuum limit, one can infer that the transition in the $\Nf=2$ chiral limit is of first order. However, should our linear extrapolation give $\NfLin \gtrsim 2$, then knowledge of the size of the scaling region is necessary to draw conclusions.

\section{Numerical strategy}
\label{sec:numericalStrategy}
We employ unimproved staggered fermions and use the RHMC algorithm~\cite{Kennedy:1998cu} to simulate any number $\Nf$ of degenerate flavors, with $\frac{\Nf}{4}$ being the power to which the fermion determinant is raised in $Z_{\Nf}(m)$.
All numerical simulations are performed using the publicly available \Ocl-based code \clqcd~\cite{Philipsen:2014mra} of which a version 1.0 has been recently released~\cite{clqcd}.
We consider temporal extents $\NTau = 4,6$ to check for the cutoff dependence of $\NfLin$.
The ranges in mass $m$ and gauge coupling constant $\beta$ of the investigated parameter space are dictated by our purpose of locating the chiral phase transition for values of the mass $m$ around the critical $\LatMassStaggeredZTwo$ value, with the temperature related to the coupling according to $T = 1/(a(\beta)\NTau)$.

To locate and identify the order of the chiral phase transition  we rely on a finite size scaling analysis of the third and fourth standardized moments of the distribution of the (approximate) order parameter.
The $\text{n}^{\text{th}}$ standardized moment for a generic observable $\mathcal{O}$ is expressed as
\begin{equation}
    B_n(\beta,m,\NSigma) = \frac{\left\langle\left(\mathcal{O} - \left\langle\mathcal{O}\right\rangle\right)^n\right\rangle}{\left\langle\left(\mathcal{O} - \left\langle\mathcal{O}\right\rangle\right)^2\right\rangle^{n/2}} \; .
\end{equation}
Being interested in the order of the thermal phase transition in the chiral limit, we consider the kurtosis $\Binder(\beta,m)$~\cite{Binder:1981sa} of the sampled $\chiralcond$ distribution, evaluated at the coupling $\beta_c$ for which $\Skewness(\beta=\beta_c,m,\NSigma)=0$, i.e.~on the phase boundary.

\begin{figure}[tp]
   \centering\includegraphics[width=1\textwidth,clip]{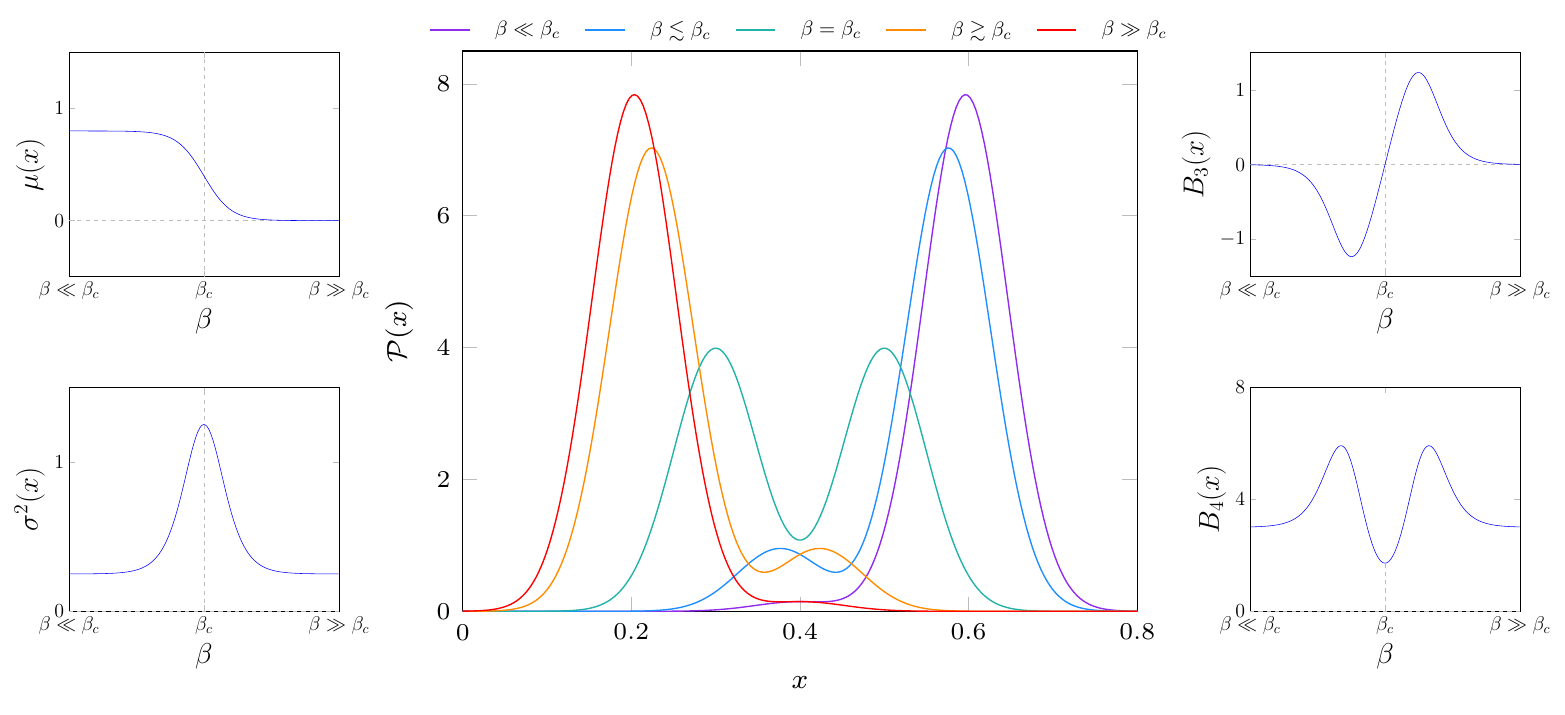}
   \caption{The chiral condensate distribution according to a model $\mathcal{P}(x)$ based on our numerical findings at various $\beta$ values and the corresponding moments as function of $\beta$. Details on the model are discussed in~\cite{Cuteri:2017gci}.}
   \label{fig:distributions}
\end{figure}

In the thermodynamic limit $\NSigma \rightarrow \infty$, the kurtosis $\Binder(\beta_c,m)$ takes the values of 1 for a first order transition and 3 for an analytic crossover, respectively, with a discontinuity when passing from a first order region to a crossover region via a second order point. For the $3D$ Ising universality class, which is the relevant one for our case, the kurtosis takes the value $1.604$~\cite{Pelissetto:2000ek}.
The discontinuous step function is smeared out to a smooth function as soon as a finite volume is considered.
In the lattice box, the distribution of the approximate order parameter and its higher moments behave, depending on $\LatCoupling$, as illustrated in Figure~\ref{fig:distributions}.
Moreover, in the vicinity of a critical point, the kurtosis $\Binder(\beta_c,m,\NSigma)$ can be expanded in powers of the scaling variable $x\equiv(m - \LatMassStaggeredZTwo) \NSigma^{1/\nu}$, and, for large enough volumes, the expansion can be truncated after the linear term,
\begin{equation}\label{eq:BinderScaling}
    \Binder(\beta_c,m, \NSigma) \simeq  \Binder(\beta_c,\LatMassStaggeredZTwo, \infty) + \textcolor{blue}{c} \, (m - \textcolor{blue}{\LatMassStaggeredZTwo}) \NSigma^{1/\nu}.
\end{equation}
As already mentioned, in our case, the critical value for the mass $\LatMassStaggeredZTwo$ is known to correspond to a second order phase transition in the 3D Ising universality class, so we fix \mbox{$\Binder(\beta_c,\LatMassStaggeredZTwo, \infty) = 1.604$} and $\nu = 0.6301$ to better constrain the fit.

Our simulated values for $\Binder(\beta_c,m, \NSigma)$ are then fitted to Eq.~\eqref{eq:BinderScaling} and the fit parameters $\textcolor{blue}{c}$ and $\textcolor{blue}{\LatMassStaggeredZTwo}$ are extracted.
The whole study has been repeated for $\Nf\in\lbrace3, 4, 5\rbrace$ at $\NTau=4$ and \mbox{$\Nf\in\lbrace3.6, 4.0, 4.4\rbrace$ at $\NTau=6$}.

\section{Results and conclusions}
\label{sec:resultsAndConclusions}

\begin{figure}[tp]
   \centering\includegraphics[width=0.9\textwidth,clip]{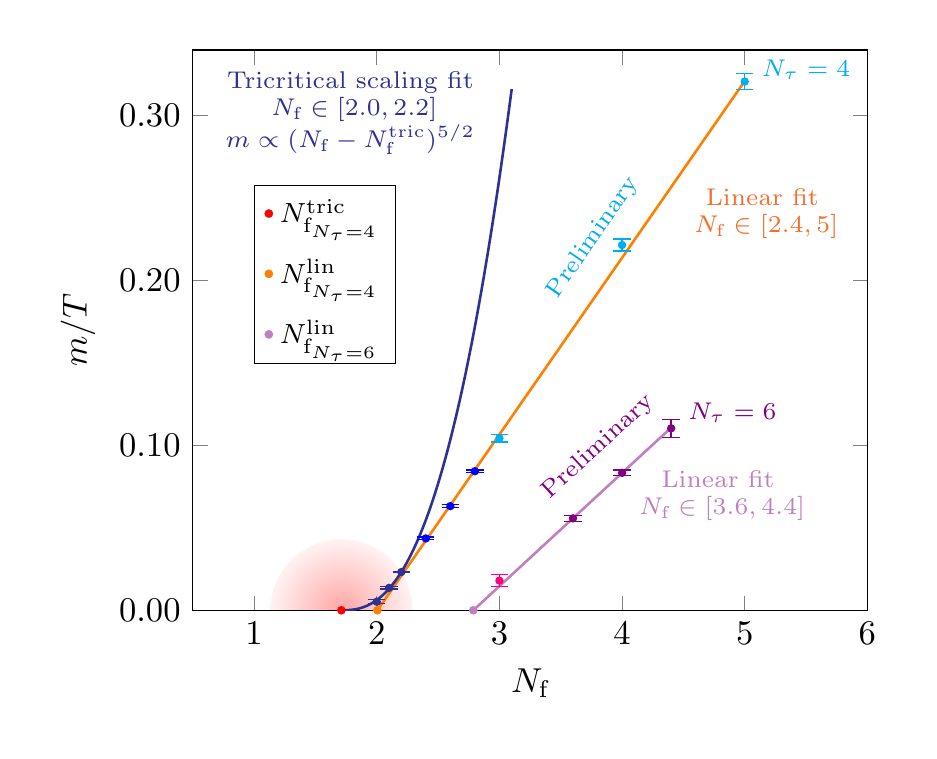}
   \caption{The $\ZTwoUniversality$ boundary in the $\LatMassStaggered/T-\Nf$ plane for $\NTau=4, 6$. The dark blue line represents the tricritical extrapolation to the chiral limit as in~\cite{Cuteri:2017gci}. The orange line represents a linear extrapolation based on $\LatMassStaggeredZTwo$ in the $\Nf$ range 2.4-5 using also newly simulated points. The violet line represents a linear extrapolation on the basis of $\LatMassStaggeredZTwo$ in the $\Nf$ range 3.6-4.4. The magenta point at $\NTau=6$ and $\Nf=3$ is borrowed from~\cite{deForcrand:2007rq}.}
   \label{fig:results}
\end{figure}

Our results are reported in Figure~\ref{fig:results}.
The first important thing to observe is that, while tricritical extrapolation for $\NTau=4$ resulted in $\NfTricr < 2$, providing a confirmation for the first order scenario being realized on coarse lattices, a linear extrapolation to the chiral limit using those data which exhibit linear scaling within the range $\Nf \in [2.4,5.0]$, results in $\NfLin=2$ within errors.
Strictly speaking, by just considering $\NTau=4$ results, one would conclude that the linear extrapolation alone cannot give conclusive answers on the order of the $\Nf=2$ transition in the chiral limit.
However, results on finer lattices were produced as well.
On $\NTau=6$, what we observe is that data within the range $\Nf \in [3.6,4.4]$ certainly do not fall in the tricritical scaling region, but they do exhibit linear scaling.
Moreover, if we consider the result for $\Nf=3$ for the same discretization from the literature~\cite{deForcrand:2007rq}, we can see it is fully consistent with our linear extrapolation.
Finally, the most important aspect of this result is that, linearly extrapolating at $\NTau=6$, we get $\NfLin \lesssim 3$, namely quite far to the right of $\Nf=2$.

To conclude, we have proposed and tested an approach, to clarify the order of the thermal transition in the chiral limit of QCD at zero chemical potential with two dynamical flavors of quarks.
Specifically, a controlled chiral extrapolation in the $m-N_f$ plane with $N_f$ promoted to a continuous parameter in the path integral formulation of the theory is possible, given that if the transition for $m\rightarrow0$ changes with $N_f$ from $1^{st}$ order (triple) to $2^{nd}$ by reducing $\Nf$, there has to exist a tricritical point at some $\NfTricr$.
Moreover, the linearity featured by the $\ZTwoUniversality$ boundary over a wide $\Nf$ region suggests that a linear extrapolation to $m=0$ can also provide an upper bound $\NfLin$ for $\NfTricr$, which may become useful to discriminate between first and second order scenario and help resolving the "$\Nf=2$ puzzle".

% On our coarse lattices the conclusion for $\Nf=2$ is that
% 
% the first order scenario seems to be realized on $\NTau=4$ lattices, in agreement with earlier results using imaginary chemical potential, instead of $\Nf$, in the same strategy.
% 
% already on $\NTau=6$ lattices the first order scenario looks more improbable.

Based on our numerical findings, the shift in the $\ZTwoUniversality$ critical boundary from $\NTau=4$ to $\NTau=6$ points towards a behavior consistent with that from improved actions on sufficiently fine lattices.

%----------------------------------------------------------------------------
\section{Acknowledgements}
The authors acknowledge support by the Deutsche Forschungsgemeinschaft (DFG) through the grant CRC-TR 211 ``Strong-interaction matter under extreme conditions'' and by the Helmholtz International Center for FAIR within the LOEWE program of the State of Hesse.
The project received initial support by the German BMBF under contract no. 05P1RFCA1/05P2015 (BMBF-FSP 202).
We also thank the staff of \Loewe\ and \Lcsc\ for their support.

\bibliographystyle{JHEP}
\bibliography{lattice2018}

\end{document}